\documentclass[letter]{aa} % for the letters 
\usepackage{graphicx}
\usepackage{txfonts}
\usepackage[]{hyperref}
\usepackage{natbib}
\usepackage{placeins}

\begin{document} 

\title{Inference of electric currents in the solar photosphere}

\author{A. Pastor Yabar\inst{1} \and 
  J.M.~Borrero\inst{2} \and 
  C. Quintero Noda\inst{3,4} \and 
  B.~Ruiz Cobo\inst{3,4}}
\institute{Institute for Solar Physics, Department of Astronomy, Stockholm University, AlbaNova University Centre, SE-10691 Stockholm, Sweden
\and
Leibniz-Institut f\"ur Sonnenphysik, Sch\"oneckstr. 6, D-79110, Freiburg, Germany
\and
Instituto de Astrof{\'{\i}}sica de Canarias, Avd. V{\'{\i}}a L\'actea s/n, E-38205, La Laguna, Tenerife, Spain
\and
Departamento de Astrof{\'{\i}}sica, Universidad de La Laguna, E-38205, La Laguna, Tenerife, Spain
}
   \date{Received September 15, 1996; accepted March 16, 1997}

\abstract{Despite their importance, routine and direct measurements of electric currents, ${\bf j}$, in the solar atmosphere have generally not been possible.}
{ We aim at demonstrating the capabilities of a newly developed method for determining electric currents in the solar photosphere.}
{We employ three-dimensional radiative magneto-hydrodynamic (MHD) simulations to produce synthetic Stokes profiles in several spectral lines with a spatial resolution similar to what the newly operational 4-meter Daniel K. Inouye Solar Telescope (DKIST) solar telescope should achieve. We apply a newly developed inversion method of the polarized radiative transfer equation with magneto-hydrostatic (MHS) constraints to infer the magnetic field vector in the three-dimensional Cartesian domain, $\mathbf{B}(x,y,z),$ from the synthetic Stokes profiles. We then apply Ampere's law to determine the electric currents, ${\bf j}$, from the inferred magnetic field, $\mathbf{B}(x,y,z),$ and compare the results with the electric currents present in the original MHD simulation.}
{We show that the method employed here is able to attain reasonable reliability (close to 50 \% of the cases are within a factor of two, and this increases to 60 \%-70 \% for pixels with $B\ge300$ G) in the inference of electric currents for low atmospheric heights (optical depths at 500 nm $\tau_{5}\in$[1,0.1]) regardless of whether a small or large number of spectral lines are inverted. Above these photospheric layers, the method's accuracy strongly deteriorates as magnetic fields become weaker and as the MHS approximation becomes less accurate. We also find that the inferred electric currents have a floor value that is related to low-magnetized plasma, where the uncertainty in the magnetic field inference prevents a sufficiently accurate determination of the spatial derivatives.}
{We present a method that allows the inference of the three components of the electric current vector at deep atmospheric layers (photospheric layers) from spectropolarimetric observations.}
\maketitle

\def\kms{~km s$^{-1}$}
\def\deg{^{\circ}}
\def\df{{\rm d}}
\newcommand{\ve}[1]{{\rm\bf {#1}}}
\newcommand{\diff}{{\rm d}}
\newcommand{\Conv}{\mathop{\scalebox{1.5}{\raisebox{-0.2ex}{$\ast$}}}}%
\def\ex{{\bf e_x}}
\def\ez{{\bf e_z}}
\def\ey{{\bf e_y}}
\def\expr{{\bf e_x^\ensuremath{\prime}}}
\def\ezpr{{\bf e_z^\ensuremath{\prime}}}
\def\eypr{{\bf e_y^\ensuremath{\prime}}}
\def\xp{x^\ensuremath{\prime}}
\def\yp{y^\ensuremath{\prime}}
\def\zp{z^\ensuremath{\prime}}
\def\rp{r^\ensuremath{\prime}}
\def\xas{x^{\ast}\!}
\def\yas{y^{\ast}\!}
\def\zas{z^{\ast}\!}
\def\C{\mathcal{C}}

%%%%%%%%%%%%%%%%%%%%%%%
\section{Introduction}
%%%%%%%%%%%%%%%%%%%%%%%
\label{sec:introduction}

Electric currents play a very important role in the energy balance and transport within the plasma in the solar atmosphere, particularly in the solar corona \citep{priest1998heating}. They are also proxies for magnetic reconnection, which is considered to be the main driver of explosive and transient events in the corona \citep{priest2002flare,amari2014flare, green2018review,mark2019flare}. Consequently, one of the main goals for the solar physics community is to measure these electric currents \citep{rolf2019est}. The most straightforward location in which to measure them is the solar photosphere. Here, electric currents are largest because of the strong braiding of the magnetic field lines caused by convective flows \citep{gudiksen2005heating}.\\

Despite their great importance, observational studies have usually neglected to infer electric currents. In order to calculate the electric currents, we must first determine the magnetic field vector, ${\bf B}$, in the whole $x,y,z$ domain and then apply Ampere's law, ${\bf j} = c (4\pi)^{-1} \nabla \times {\bf B}$ \footnote{In the solar atmosphere it is possible to neglect the time variations of the electric field, and thus Maxwell's equation becomes Ampere's law \citep{priest1984}}. The inference of the magnetic field vector, {\bf B}, is mostly done via Zeeman polarimetry (by means of spectropolarimetry) and the application of inversion techniques to the polarized radiative transfer equation \citep[RTE;][]{jc2016review}.\\

Traditional Milne-Eddington (ME) inversion techniques \citep{borrero2011vfisv,borrero2014milne} infer a magnetic field that is constant with optical depth. Thus, one retrieves an average magnetic field over the optical depth region where the spectral lines are formed, ${\bf B}(x,y,\tau_c^{*})$, where $\tau_c^{*}$ represents a ``characteristic'' optical depth \cite[typically around where the considered spectral line shows the largest response;][]{westendorpplaza1998}. However, even if the corrugation in the $z$ scale of the $\tau_c^{*}$ isosurfaces (i.e., the fact that light coming from different atmospheric regions might be coming from different solar atmosphere heights; see, for example, the Wilson effect) is ignored, from ME inversions it is only possible to determine the vertical component of the electric current, $j_z(x,y,\tau_c^{*})$. This is so because, by hypothesis, ME (among other premises) assumes no variation with $z$ in the magnetic field vector. Because of this, many studies have been carried out where only one component of the electric current is considered \citep{pevtsov1990currents, metcalf1994, solanki2003he, wang2017flare}.\\

More sophisticated inversion techniques \citep{basilio1992sir, frutiger2000spinor, delacruzrodriguez2019} allow ${\bf B}$  to be inferred in the three-dimensional $(x,y,\tau_c)$ domain (i.e., allow the magnetic field to also be retrieved as a function of optical depth instead of a fixed value and so allow the computation of the x and y components of $\mathbf{j}$). From there, it is possible to determine the full ${\bf j}$ vector if we can first determine the $z$ scale. This is typically done via the assumption of hydrostatic equilibrium \citep[HE; see, for instance,][]{hector2005currents}. However, as discussed in \cite{borrero2019mhs}, this method yields a very poor determination of the $z$ scale in the presence of magnetic fields.

\begin{table*}
\caption{Spectral lines and their associated atomic parameters for the three spectral ranges considered.}              % title of Table
\label{tab:spectral_ranges}      % is used to refer this table in the text
\centering                                      % used for centering table
\begin{tabular}{c c c c c c c c c c c c c}          % centered columns (4 columns)
\hline\hline                        % inserts double horizontal lines
Spectral range & $\Delta\lambda$ & ${\rm n}_{\lambda}$ & $\lambda_{0}$ & ${\rm e}_{\rm low}$ & ${\rm e}_{\rm upp}$ & $\log_{10}{{\rm g}\,{\rm f}}$ & ${\rm E}_{\rm low}$ & $\alpha$ & $\sigma/{\rm a}_{0}^{2}$ \\    % table heading
 & [m{\AA}] &  & [{\AA}] &  &  &  & [eV] &  &  &  \\
\hline                                   % inserts single horizontal line    Fe\,{\sc i} 630.2 nm   & 20 & 112 & 6301.501 & $^{5.0}{\rm P}_{2.0}$ & $^{5.0}{\rm D}_{2.0}$ & -0.718 & 3.654 & 0.243 & 835.356\\      % inserting body of the table
                            &       & \tablefootmark{a} & 6302.494 & $^{5.0}{\rm P}_{1.0}$ & $^{5.0}{\rm D}_{0.0}$ & -1.236 & 3.686 & 0.239 & 850.930\\      % inserting body of the table
    Si\,{\sc i} 1082.7 nm    & 18    & 278 & 10827.091 & $^{3.0}{\rm P}_{2.0}$ & $^{3.0}{\rm P}_{2.0}$ & 0.239 & 4.953 & 0.231 & 729.807\\
    Fe\,{\sc i} 1565.0 nm & 40    & 125 & 15648.515 & $^{7.0}{\rm D}_{1.0}$ & $^{7.0}{\rm D}_{1.0}$ & -0.669 & 5.426 & 0.229 & 974.195\\
                            & 40    & 125 & 15662.018 & $^{5.0}{\rm F}_{5.0}$ & $^{5.0}{\rm F}_{4.0}$ & 0.19 & 5.83 & 0.240 & 1196.950\\
\hline                                             %inserts single line
\end{tabular}
\tablefoot{The first column is the label used for each spectral region (which might include more than one spectral line). In the other columns, $\Delta\lambda$ is the pixel size in m{\AA}; ${\rm n}_{\lambda}$ is the number of wavelengths used for that spectral line; $\lambda_{0}$ is the central wavelength for the electronic transition associated with the spectral line; ${\rm e}_{\rm low}$ and ${\rm e}_{\rm upp}$ are the electronic configurations of the lower and upper energy level, respectively; ${\rm E}_{\rm low}$ is the excitation potential (in eV) of the lower energy level; and $\alpha$ and $\sigma/{\rm a}_{0}^{2}$ are the velocity exponent and collision cross section parameters, respectively, as defined in Anstee, Barklem, and O'Mara collision theory for the broadening of metallic lines by neutral hydrogen collisions \citep{anstee1995, barklem1998}.\\
\tablefoottext{a}{This spectral line is considered a blend with the spectral line with $\lambda_{0}=6301.501$ {\AA}.}
}
\end{table*}

Although there are a number of methods available to infer the physical parameters in the ($x,y,z$) domain \citep{puschmann2010pen, riethmuller2017, loeptien2018zw, andres2019invz}, to our knowledge only the method presented in \cite{puschmann2010pen} has been employed in the determination of the full electric current vector, ${\bf j}$, although the authors did not address the reliability to which electric currents can be determined. Motivated by their results, we have developed a new inversion code for the RTE -- the forward-inverse solver of the polarized RTE under the Zeeman regime in geometrical scale, \texttt{FIRTEZ-dz} -- that retrieves the magnetic field, $ {\bf B}$, in the $(x,y,z)$ domain instead of the $(x,y,\tau_c)$ domain \citep{adur2019invz,borrero2019mhs,borrero2021} via the application of magneto-hydrostatic (MHS) equilibrium instead of HE. In the present Letter we demonstrate that our method can also be employed to determine the three components of the electric current, ${\bf j}$.

%%%%%%%%%%%%%%%%%%%%%%%
\section{Synthetic observation and inference methods}
%%%%%%%%%%%%%%%%%%%%%%%

In order to address the reliability of the determination of electric currents, we made use of a state-of-the-art radiative magneto-hydrodynamic (MHD) simulation snapshot of a sunspot \citep{rempel2012mhd}. The dimension of the simulation box is 4096$\times$4096$\times$768 pixels with a grid size of 12, 12, and 8 km, respectively, with the third coordinate directed along the direction of gravity. Of the whole spatial domain, we focused on a small region of 512$\times$512 pixels close to the main sunspot, in which there are two small pores surrounded by quiet Sun (the pores are highlighted in panel a of Fig. \ref{fig:spatial_distribution} with a cyan contour, and some atmospheric properties are shown in Appendix \ref{app:inversionresults}). As for height, we considered only the uppermost 192 grid cells. This is sufficient to account for the entire photosphere (plus the lower chromosphere) in both the quiet Sun and the pore, including its Wilson depression. The electric currents (in Gaussian units) are calculated following ${\bf j}=\frac{c}{4\,\pi}\nabla \times {\bf B}$, for which the Cartesian derivatives of the components of the magnetic field are required, and where $c$ is the speed of light in a vacuum. In the following, the electric currents thus derived from the magnetic field of the MHD simulation (${\bf j}_{\rm MHD}$) will remain as the underlying truth against which we compare the derived electric currents from the inversion of synthetic observations (${\bf j}_{\rm inv}$). To do so, it is important to note that the calculation of the electric currents themselves is performed in the (x,y,z) domain, but the results will be shown at different $\tau_{5}$ isosurfaces. This is done because the geometrical heights at which our selected spectral regions (see below) are sensitive depend on the physical atmospheric properties (mostly on the interplay between the temperature and gas pressure with height) at a given $(x,y)$ point on the solar surface.

\begin{figure*}
\centering
   \includegraphics[width=17cm]{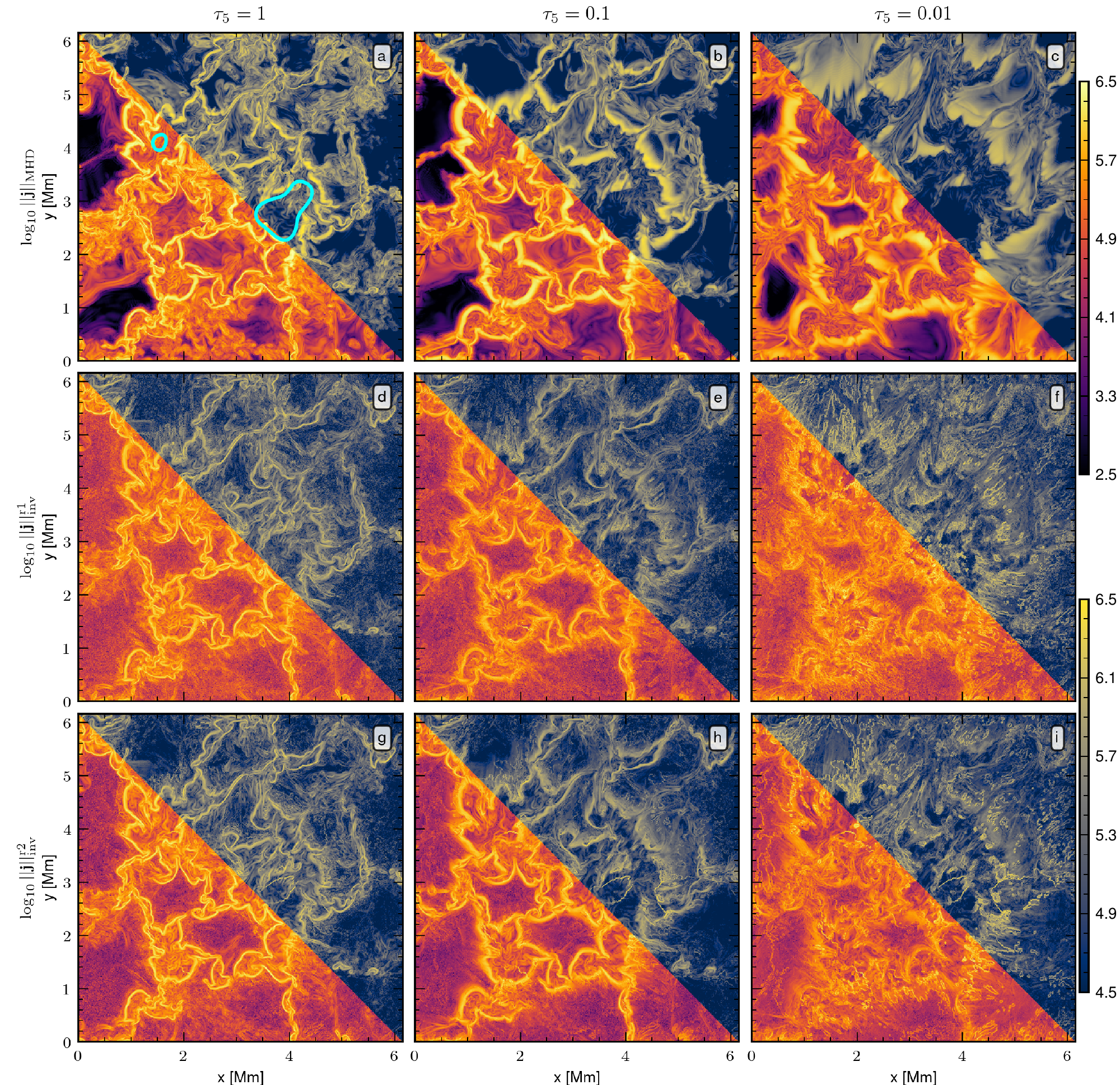}
   \caption{Electric currents, ${\bf j}$ (in decimal logarithm of the L2 norm of ${\bf j}$ in ${\rm g}^{1/2}\,{\rm cm}^{-1/2}\,{\rm s}^{-2}$), and spatial distribution for three $\tau_{5}$ isosurfaces (from left to right: $\tau_{5}=1$, $\tau_{5}=0.1$, and $\tau_{5}=0.01$). From top to bottom: Electric currents for the MHD snapshot (${\bf j}_{\rm MHD}$), for the inversion of the 6301/2{\AA} spectral range (${\bf j}_{\rm inv}^{\rm r1}$), and for the inversion of all the spectral ranges gathered in Table \ref{tab:spectral_ranges} (${\bf j}_{\rm inv}^{\rm r2}$). Two color maps are used: one includes all the wealth of the simulation features (the reddish one), and the other focuses on the electric currents above the floor value and allows an easier comparison to ${\bf j}_{\rm MHD}$ (the blueish one). The cyan contour in panel a represents the normalized continuum intensity threshold of 0.65, highlighting the position of the pores.}
     \label{fig:spatial_distribution}
\end{figure*}

We used three different spectral regions: Fe\,{\sc i} 630.2 nm, Si\,{\sc i} 1082.7 nm, and Fe\,{\sc i} 1565.0 nm. The atomic data and spectral setup employed are presented in Table \ref{tab:spectral_ranges}. The synthesis of the four Stokes parameters is performed using the local thermodynamic equilibrium (LTE) RTE solver \texttt{FIRTEZ-dz} \citep{adur2019invz}\footnote{The silicon spectral line considered here is known to present non-LTE effects close to its core \citep{shchukina2017}, but since we use the same setup for the synthesis and inversion process, it has no impact in this work.}. In order to avoid possible uncertainty sources and address the determination of electric currents following the method presented in this work, we have considered the most ideal case in which we take neither the spatial point-spread function (PSF) nor the spectral line-spread function (LSF) of the hypothetical instrument into account. Here we implicitly assume that the effect of the PSF (as the LSF has commonly been taken into account) can be properly accounted for during the post-processing of the data \citep[see, for instance,][]{lofdahl2021,grinonmarin2021} or in the modeling \citep{vannoort2012decon, delacruzrodriguez2019b}. A discussion of the various approaches to handling both effects is available in \cite{delacruzrodriguez2017}. We do include normally distributed noise on the order of $10^{-3}$ for all the Stokes parameters. In our analysis, we use two sets of different synthetic observations given by a different combination of the abovementioned spectral ranges. In the first case, ${\bf S}^{\rm r1}$, we consider a single spectral range (630.2 nm). This example considers a similar spectral configuration as the spectropolarimeter \citep[SP;][]{lites2013} at the Solar Optical Telescope \citep[SOT;][]{kosugi2007,tsuneta2008a} on board the Hinode Satellite\footnote{We refer here only to a spectral sampling similar to SP as its spatial resolution is worse than the one considered in this work.}. In the second case, ${\bf S}^{\rm r2}$, we consider a multiwavelength observation in which we have access to all the spectral regions in Table \ref{tab:spectral_ranges}. This configuration maximizes the height range at which the inversion becomes sensitive, improving the inference of $\mathbf{B}(z)$, which is key in the determination of $\mathbf{j}_{x,y}$. Simultaneous observations in these three ranges can be attained, for instance, by the incoming Diffraction-Limited Near-Infrared Spectro-Polarimeter (DL-NIRSP) at the Daniel K. Inouye Solar Telescope \citep[DKIST;][]{rimmele2020} or, for an equivalent combination of spectral lines, with the Visible Spectro-Polarimeter (ViSP).

In order to infer the magnetic field, ${\bf B}$, in the ($x$, $y$, $z$) domain, we proceeded in a similar fashion as in \cite{borrero2021}, with some critical changes to both the inversion process and the solution of the MHS equation, as detailed in Appendix \ref{app:inversion}. Once the inversion is completed, we obtain the various atmospheric parameters for each run  -- in particular for this work, ${\bf B}(x,y,z)$  -- thus allowing us to compute the electric currents (see Appendix \ref{app:inversionresults} for a general overview). We performed this process for both data sets (${\bf S}^{\rm r1}$ and ${\bf S}^{\rm r2}$), thereby retrieving two different electric current vectors in the ($x$, $y$, $z$) volume (${\bf j}_{\rm inv}^{\rm r1}$ and ${\bf j}_{\rm inv}^{\rm r2}$, respectively).

%%%%%%%%%%%%%%%%%%%%%%%
\section{Results}
\label{sec:results}
%%%%%%%%%%%%%%%%%%%%%%%

In Fig. \ref{fig:spatial_distribution} we explore the spatial distribution of the electric currents inferred at different optical depths as determined from the magnetic field coming from the MHD simulation (top row), the inversion of ${\bf S}^{\rm r1}$ (middle), and the inversion of ${\bf S}^{\rm r2}$ (bottom). As can be seen, at $\tau_{5}=1$ and $\tau_{5}=0.1$, ${\bf j}_{\rm inv}^{\rm r1}$ and ${\bf j}_{\rm inv}^{\rm r2}$ (panels d-g and e-h, respectively) closely resemble the main small-scale structures as well as the general spatial distribution of ${\bf j}_{\rm MHD}$ (panels a-b). The agreement between ${\bf j}_{\rm MHD}$ and the inferred ones (${\bf j}_{\rm inv}^{\rm r1}$ and ${\bf j}_{\rm inv}^{\rm r2}$) clearly worsens at $\tau_{5}=0.01$ in spite of the inclusion of the Si\,{\sc i} 1082.7 nm spectral line that conveys information from higher atmospheric layers as compared to the other two spectral regions in Table \ref{tab:spectral_ranges}. A striking feature worth noting is that the lowest values of ${\bf j}_{\rm MHD}\approx10^{2.5-4.}$ (darkish regions in panels a-c) are not properly inferred by our method. Instead, our inversions yield a floor value of ${\bf j}_{\rm inv}\approx10^{4}$. We refer to this floor value, below which we cannot infer reliable electric currents, as ${\bf j}_{\rm inv, min}$. Interestingly, the floor value for the electric currents is larger for the inversion of ${\bf S}^{\rm r1}$ than for ${\bf S}^{\rm r2}$, that is to say, the inclusion of more spectral lines allows a smaller lower electric current determination: ${\bf j}_{\rm inv, min}^{\rm r1}\approx10^{4.7}\,{\rm g}^{1/2}\,{\rm cm}^{-1/2}\,{\rm s}^{-2}$ as compared to ${\bf j}_{\rm inv, min}^{\rm r2}\approx10^{4.4}\,{\rm g}^{1/2}\,{\rm cm}^{-1/2}\,{\rm s}^{-2}$.

\begin{figure*}
\centering
  \includegraphics[width=\textwidth]{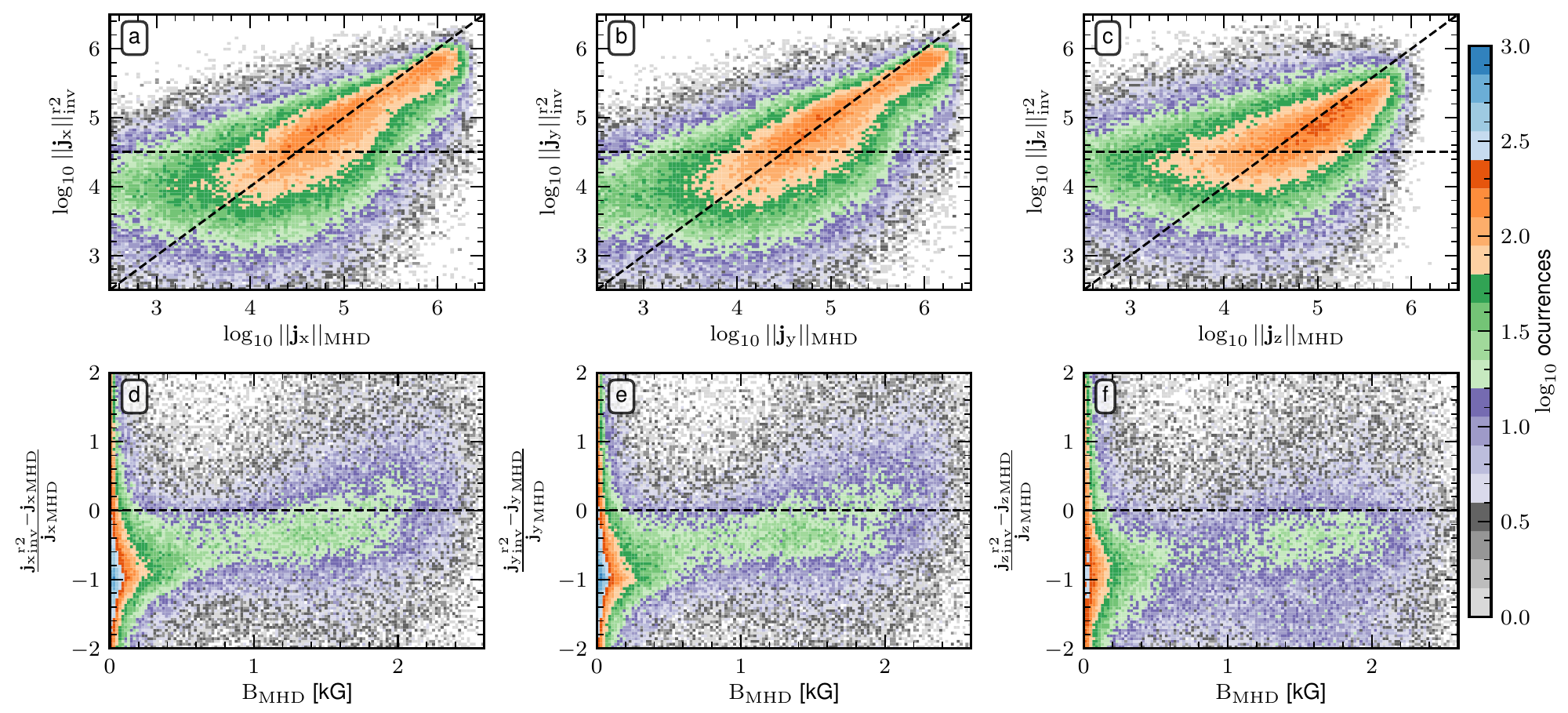}
     \caption{Electric current components for the second inversion setup (${\bf j}_{\rm inv}^{\rm r2}$) as compared to the MHD case for the $\tau_{5}=0.1$ isosurface. Panels a, b, and c show the bidimensional histogram of each electric current vector component ($x$, $y$, $z$) as compared to the MHD ones (${\bf j}_{\rm MHD}$). The horizontal black-dashed line highlights the floor electric current value found for $||{\bf j}||_{\rm MHD}>10^{4.4}\,{\rm g}^{1/2}\,{\rm cm}^{-1/2}\,{\rm s}^{-2}$ and the oblique one the one-to-one relation. Panels d, e, and f show the dependence with the MHD magnetic field strength of the relative error of each electric current vector component  ($x$, $y$, $z$) as compared to the MHD cases (${\bf j}_{\rm MHD}$). The horizontal black-dashed line highlights the equality line. The color code is the decimal logarithm of the occurrences.}
     \label{fig:j_components}
\end{figure*}

A more quantitative study of the accuracy to which we can determine electric currents can be done by finding the percentage of pixels where the inferred electric currents, ${\bf j}_{\rm inv}^{\rm r1}$, are within a factor of two of the original currents in the MHD simulations, ${\bf j}_{\rm MHD}$. The numbers for $\tau_{5}=1$, $\tau_{5}=0.1$, and $\tau_{5}=0.01$ are 46.03 \%, 40.96 \%, and, 33.80 \%, respectively. In a wider view, the percentage of pixels in which our method retrieves the electric currents within an order of magnitude (factor of ten) of ${\bf j}_{\rm MHD}$ are 88.51 \%, 83.93 \%, and 85.42 \% for the abovementioned optical depths. For the second run, ${\bf S}^{\rm r2}$ , the fraction of pixels with ${\bf j}_{\rm inv}^{\rm r2}$ within a factor of two of the ${\bf j}_{\rm MHD}$ are 49.23 \%, 46.24 \%, and 35.33 \% for $\tau_{5}=1$, $\tau_{5}=0.1$, and $\tau_{5}=0.01$, respectively, and 90.05 \%, 86.18 \%, and 88.60 \% of ${\bf j}_{\rm inv}^{\rm r2}$ are inferred within an order of magnitude as compared to ${\bf j}_{\rm MHD}$. Another common result for both observational setups (${\bf S}^{\rm r1}$ and ${\bf S}^{\rm r2}$) is that the inferred electric currents are predominantly underestimated. Thus, for $\approx$ 60 \% of the pixels, the inferred currents are underestimated as compared to the MHD one, regardless of being within a factor of two or ten of the MHD values.

In the following we further explore the source for the floor value on the inferred electric current, ${\bf j}_{\rm inv, min}$. To this end, we include in Fig. \ref{fig:j_components} the Cartesian components of ${\bf j}_{\rm inv}^{\rm r2}$ at $\tau_{5}=0.1$ as compared to the same components for the MHD (top row). The comparison of the $z$ component of the electric current (panel c) to the other two components of ${\bf j}_{\rm inv}^{\rm r2}$ (panels a and b) shows that, even though all three components are affected to some degree by a minimum value below which inferences are unreliable, the largest contribution to the floor value comes from the ${j_{z}}$ component of the electric current. This is seen as a horizontal distribution around ${j_{z}}_{\,\rm inv}^{\rm r2}\approx10^{4.4}\,{\rm g}^{1/2}\,{\rm cm}^{-1/2}\,{\rm s}^{-2}$ for ${j_{z}}_{\,\rm MHD}<10^{4.4}\,{\rm g}^{1/2}\,{\rm cm}^{-1/2}\,{\rm s}^{-2}$ in panel c as compared to similar ones appearing at ${j_{x(y)}}_{\,\rm inv}^{\rm r2}\approx10^{3.9}\,{\rm g}^{1/2}\,{\rm cm}^{-1/2}\,{\rm s}^{-2}$ for $\approx{j_{x(y)}}_{\,\rm MHD}<10^{3.9}\,{\rm g}^{1/2}\,{\rm cm}^{-1/2}\,{\rm s}^{-2}$.

The $j_{z}$ is related to the $x$ and $y$ derivatives of the $y$ and $x$ components, respectively, of the magnetic field. With this in mind, we surmise that the likely reason as to why the floor value in the determination of the electric currents is dominated by $j_{z}$ instead of $j_{x}$ or $j_{y}$ is the lower sensitivity to linear polarization (which depends on $B_{x}$ and $B_{y}$) as compared to circular polarization, and so it is expected that $B_{x}$ and $B_{y}$ are less accurate for low polarimetric signals. It could be argued that both $j_{x}$ and $j_{y}$  include vertical $z$ derivatives of $B_{x}$ and $B_{y}$, and therefore there should be no reason as to why these two components should be better retrieved than $j_{z}$. However, we note here that the $z$ derivatives of $B_{x}$ and $B_{y}$ are regularized via Tikhonov’s method, while this is not the case (so far in our method) for the $x$ and $y$ derivatives. Moreover,  $j_{x}$ and $j_{y}$ -- the $y$ and $x$ derivatives, respectively, of $B_{z}$ -- are also included, and they are extremely well retrieved owing to the larger sensitivity to circular polarization. Under these circumstances one would expect the accuracy in the determination of electric currents to improve as the magnetic field increases as it should allow a better characterization of the global behavior of the magnetic field, if not the full complexity present in the MHD simulation. In order to test this hypothesis, we include in the bottom row of Fig. 2 (panels d, e, and f) the variation in the normalized comparison of each inferred and MHD electric current component with the magnetic field strength (as given in the MHD). As is clear for all three components, as the magnetic field increases, the relative error in each electric current becomes closer to zero. As an example, if we calculate how the inferred electric currents compare to the MHD ones (as we did above), but now considering only pixels with magnetic field strength $B\ge300$ G, then we get that the inferred electric currents, ${\bf j}_{\rm inv}^{\rm r1}$, are within a factor of two from MHD ones in 58.97 \%, 57.36 \%, and 36.40 \% of the cases for $\tau_{5}=1$, $\tau_{5}=0.1$, and $\tau_{5}=0.01$, respectively, and in 68.22 \%, 67.74 \%, and 40.89 \% for the second run, ${\bf S}^{\rm r2}$.

\section{Discussion and conclusion}

In this Letter we have demonstrated that a good estimation of the electric current vector (at photospheric layers) is now possible thanks to new advances in the inversion process for the RTE coupled with MHS constraints. This is a major leap forward in our understanding of the energy balance of the solar atmosphere as well as in the understanding of the magnetic topology preceding magnetic reconnection events. This is possible thanks to the new approach explored in \cite{borrero2021}, in which the authors exploit the possibility of iteratively inverting Stokes spectropolarimetric data in a three-dimensional ($x$, $y$, $z$) volume via the reevaluation of the gas pressure that results from the solution of the MHS equation of motion. The application of Tikhonov’s regularization \citep{delacruzrodriguez2019} to better constrain the vertical derivatives of the physical parameters during the inversion process also plays a very important role in allowing for a reliable determination of the electric currents.

In order to prove this capability, we have considered an ideal case (with neither spatial nor spectral instrumental degradation) in which the Stokes vectors from a different number of spectral lines, probing different atmospheric layers, are recorded. We have demonstrated that it is possible to infer within a factor of two the real values present in the MHD simulation, ${\bf j}_{\rm MHD}$, on close to 50 \% of the solar surface. For regions where B $\ge$ 300 G, the fraction of pixels within a factor of two of ${\bf j}_{\rm MHD}$ increases to approximately 60 \% for ${\bf S}^{\rm r1}$ and 70\% for ${\bf S}^{\rm r2}$. These results are valid for an optical depth range between $\tau_{5}\in$[1,0.1]. We have found that ${\bf j}_{\rm inv}$ values suffer from a floor value of electric currents below which inferences of electric currents are not reliable. This floor value, ${\bf j}_{\rm inv, min}$, is due to the limitation in the determination of $j_{\rm z}$, which in turn is due to the fact that the perpendicular components of the magnetic field vector ($B_{\rm x}$ and $B_{\rm y}$) are sensitive to Stokes Q and U signals (i.e., linear polarization), whose amplitude is of second order as compared to Stokes V (i.e., circular polarization).

Consequently, we have determined that the accuracy in the determination of $j_{\rm x}$ and $j_{\rm y}$ is significantly higher than for $j_{\rm z}$. Interestingly, $j_{\rm z}$ is the component of the electric current, ${\bf j}$, that has been studied most often as it is the only one that does not depend on the $z$ variation of the magnetic field components and therefore can be calculated through traditional ME inversion techniques \citep{borrero2014milne, wang2017flare}. Thus, the method presented in this work has the potential to significantly deepen our understanding of the energy balance in the solar atmosphere as it allows all three components of the electric current, ${\bf j}$, to be inferred.

We have also found that the inclusion of spectral lines that probe higher layers does not significantly improve the inference for $\tau_{\rm 5}\le 0.01$. We can offer different explanations as to why this is the case. On the one hand, the MHS approximation is less accurate at higher layers as the advection terms in the equation of motion become more relevant. On the other hand, the fact that the magnetic field intensity decreases with height makes it more difficult to correctly estimate its components and consequently their spatial derivatives (i.e., electric currents).

The spatial sampling considered here is achievable by the new generation of 4-meter solar telescopes: DKIST and the European Solar Telescope \citep[EST;][]{collados2013}. In a future work, we will assess to what extent spatial degradation affects the reliability of the inferred electric currents, as well as the potential benefit that horizontal spatial regularization \citep{delacruzrodriguez2019b, morosin2020} might have in alleviating the significance of the electric current floor value. It is also important to explore the effect of the noise level (here $SNR=1000$) both on the ability of inferring electric currents at high layers ($\tau_{5}\le0.01$) and on the reliability of their determination for lowly magnetized areas.

\begin{acknowledgements}
This work has received funding from the European Research Council (ERC) under the European Union's Horizon 2020 research and innovation programme (SUNMAG, grant agreement 759548) and from the Deutsche Forschungsgemeinschaft (DFG project number 321818926). The Institute for Solar Physics is supported by a grant for research infrastructures of national importance from the Swedish Research Council (registration number 2017-00625). JMB acknowledges travel support from the Spanish Ministry of Economy and Competitiveness (MINECO) under the 2015 Severo Ochoa Program MINECO SEV-2015-0548 and from the SOLARNET project that has received funding from the European Union’s Horizon 2020 research and innovation programme under grant agreement no 824135. CQN was supported by the EST Project Office, funded by the Canary Islands Government (file SD 17/01) under a direct grant awarded to the IAC on ground of public interest, and this activity has also received funding from the European Union's Horizon 2020 research and innovation programme under grant agreement No 739500. This research has made use of NASA's Astrophysics Data System. We acknowledge the community effort devoted to the development of the following open-source packages that were used in this work: \texttt{numpy} \citep[numpy.org][]{numpy2020}, \texttt{matplotlib} \citep[matplotlib.org][]{hunter2007}. 
\end{acknowledgements}

\bibliographystyle{aa}
\bibliography{ms_pastoryabar}

\begin{appendix}

\section{Inversion results}
\label{app:inversionresults}

\begin{figure*}
\centering
   \includegraphics[width=16.6cm]{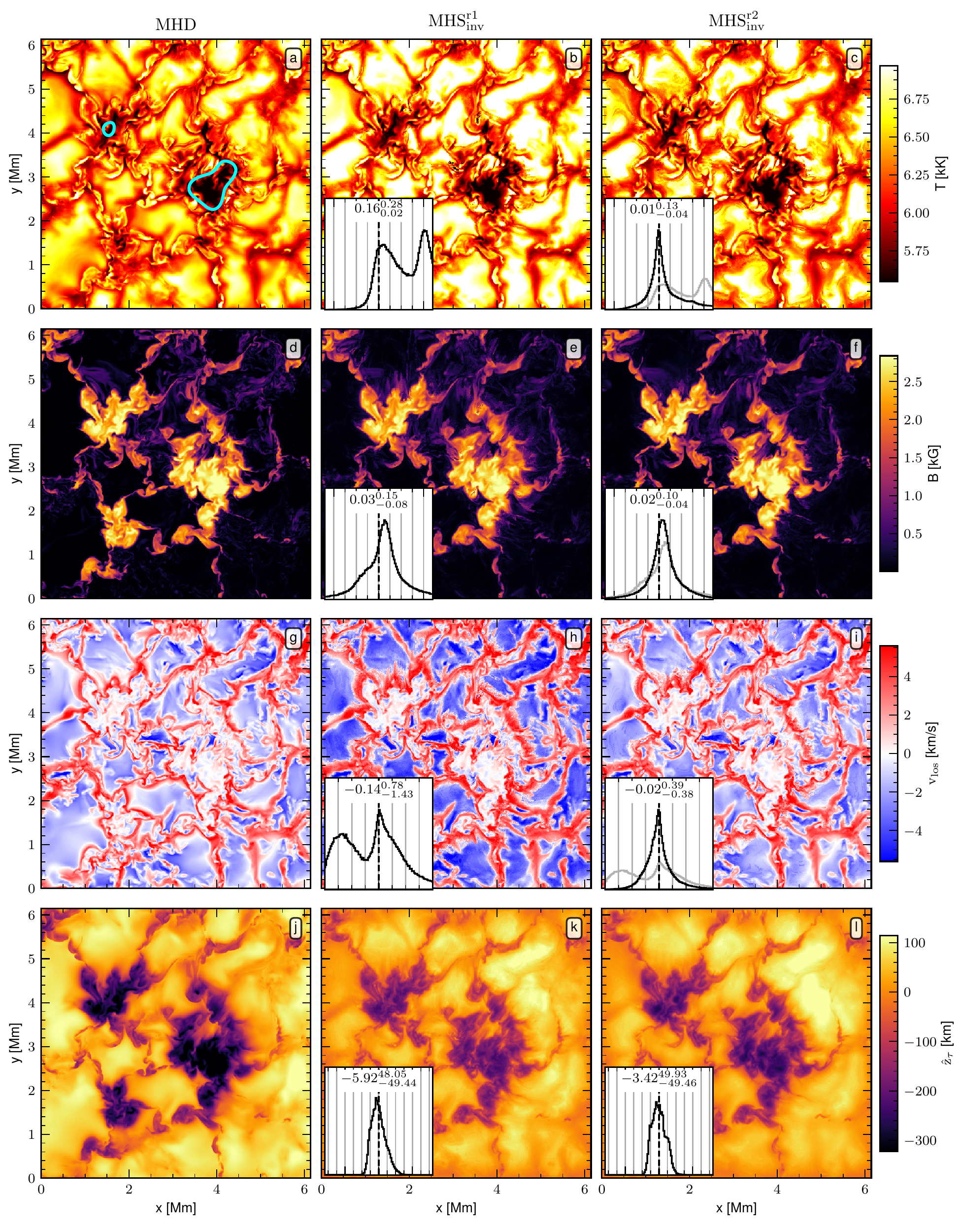}
   \caption{Temperature in kK (top row), magnetic field strength in kG (second row), line-of-sight velocity in km/s (third row), and the Wilson effect (as compared to the quiet Sun $\tau=1$ height) in km (last row) at optical depth $\tau_{\rm 5 nm}=1$ for the MHD simulation (left column), the first run of the MHS inversion (${\bf S}^{\rm r1}$, middle column), and the second run of the MHS inversion (${\bf S}^{\rm r2}$, right column). All the panels per row share the same color scaling, shown at the right. Sub-panels in panels b-c, e-f, h-i, and k-l show in black the histogram of the difference between the MHD and the MHS inversion of ${\bf S}^{\rm r1}$ (${\bf S}^{\rm r2}$) in panels b, e, h, and k (in panels c, f, i, and l). For the sake of comparison, panels c, f, i, and l show the same histograms presented in b, e, h, and k in gray. The values in the top part of each histogram are the median value together with percentile 16 as sub-index and percentile 84 as super-index. Vertical gray lines appear every 0.05 kK, 0.05 kG, 0.4 km/s, and 50 km for the temperature, magnetic field strength, line-of-sight velocity, and Wilson effect, respectively, and are centered at 0 (shown with a dark dashed line). The blue contour in panel a represents the normalized continuum intensity threshold of 0.65, highlighting the position of the pores.}
     \label{fig:inversion_results1}
\end{figure*}

\begin{figure*}
\centering
   \includegraphics[width=16.6cm]{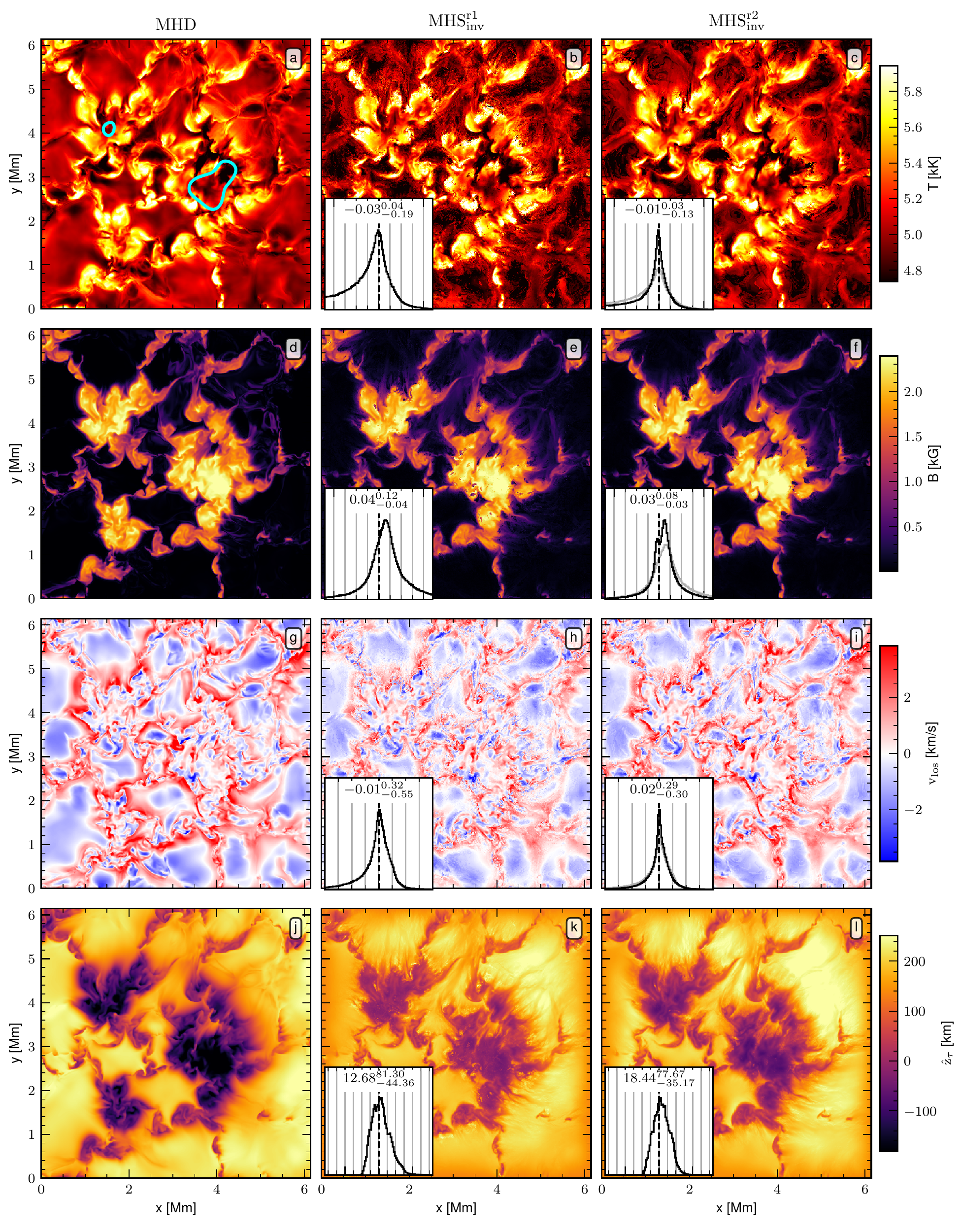}
   \caption{Same as in Fig. \ref{fig:inversion_results1} but for optical depth $\tau_{\rm 5 nm}=0.1$.}
     \label{fig:inversion_results2}
\end{figure*}

\begin{figure*}
\centering
   \includegraphics[width=16.6cm]{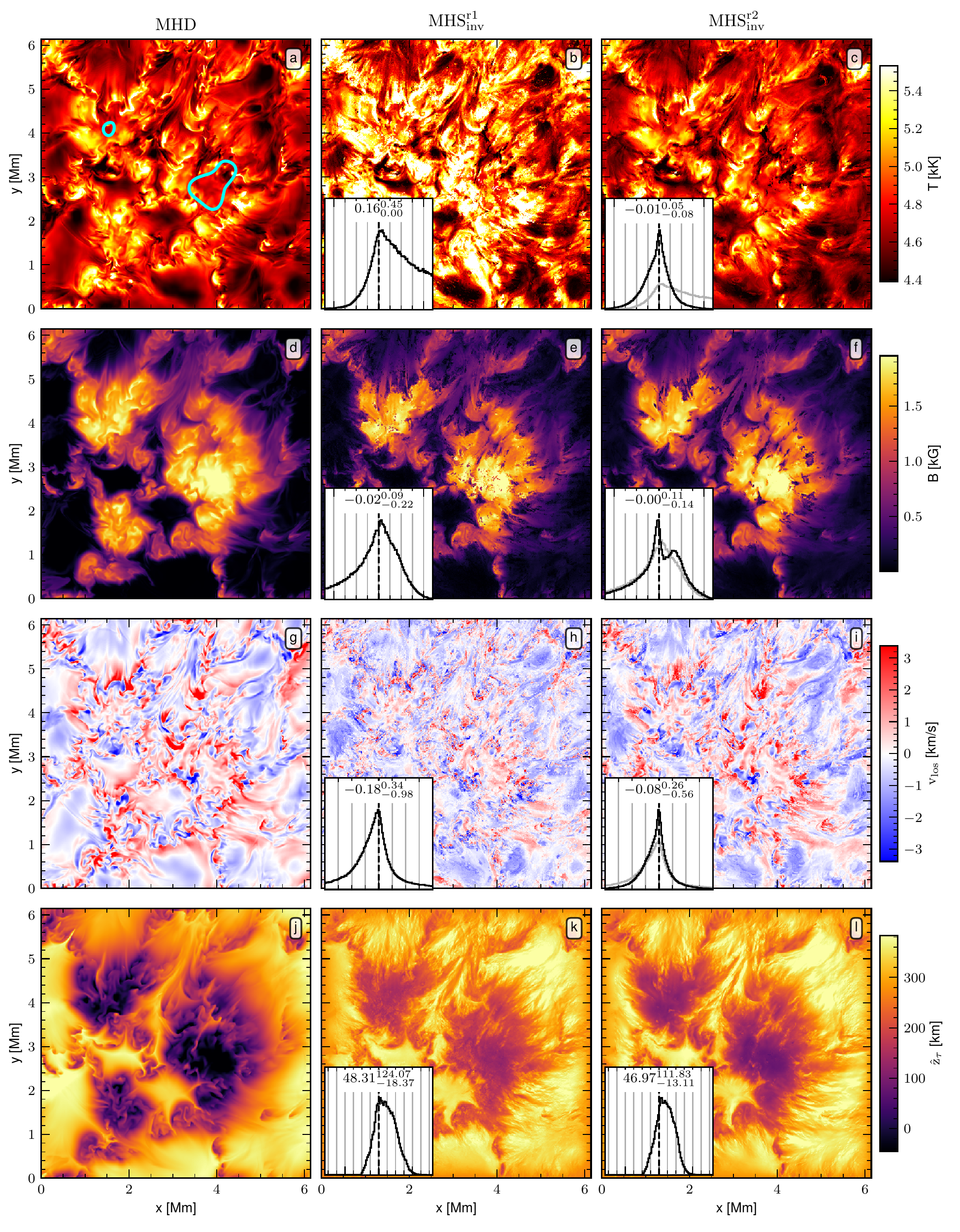}
   \caption{Same as in Fig. \ref{fig:inversion_results1} but for optical depth $\tau_{\rm 5 nm}=0.01$.}
     \label{fig:inversion_results3}
\end{figure*}

In Figs. \ref{fig:inversion_results1}, \ref{fig:inversion_results2}, and \ref{fig:inversion_results3} we present some MHD atmospheric parameter maps (temperature, magnetic field strength, line-of-sight velocity, and Wilson effect) as well as the inversion results for the three optical depths analyzed in Sect. \ref{sec:results} ($\tau_{\rm500 nm}=1$, $\tau_{\rm500 nm}=0.1$, and $\tau_{\rm500 nm}=0.01$). It should be noticed that the Wilson effect is shown relative to the height at which the surrounding quiet Sun reaches (on average) $\tau=1$, $\hat{\rm z}(\tau_{i})={\rm z}(\tau_{i})-{\rm z_{QS}(\tau=1)}$ as our method's absolute reference depends on the boundary conditions (see \citealt{borrero2019mhs} for more details). The region of interest is centered on three main strong magnetic structures (see, for instance, panel d of Fig. \ref{fig:inversion_results1}) that are clearly associated with cold plasma (see panel a), at least  for the two pores highlighted in cyan, and are surrounded by convective cells (see panels a and g). These strong magnetic features show a Wilson effect of around 300 km as compared to their neighboring quiet Sun at an optical depth unity (panel j in Fig. \ref{fig:inversion_results1}). Regarding the inversion results, there is an overall good agreement with the underlying ``truth'' (i.e., the MHD result). It is worth mentioning that while the inversion with all the spectral regions (${\rm MHS}_{\rm inv}^{\rm r2}$) achieves similar results at all three optical depths, the one with a single spectral region (${\rm MHS}_{\rm inv}^{\rm r1}$) only gets similar results (as compared to ${\rm MHS}_{\rm inv}^{\rm r2}$) for $\tau_{\rm500 nm}=0.1$. This is due to the fact that we have chosen the spectral region in ${\rm MHS}_{\rm inv}^{\rm r2}$ such that it reaches a more extended sensitivity in height thanks to the inclusion of the 1.5 $\mu$m lines that are sensitive to deep layers (see, for instance, Fig. 6 in \citealt{grinonmarin2021}) and due to the inclusion of the Si {\sc i} at $\approx$1 $\mu$m that reaches higher layers than the pair of Fe {\sc i} at $\approx$600nm.

\FloatBarrier

\section{The inversion process}
\label{app:inversion}

The inversion of the spectropolarimetric data performed in this work is based on the one presented in \cite{borrero2021}, and the reader is referred to that work for a full in-depth analysis of the method itself. It is to be noted, though, that we included two changes to the code, in the inversion process and in the solution of the MHS equation. In the former case, we included the Tikhonov regularization in \texttt{FIRTEZ-dz,} as in \cite{delacruzrodriguez2019} (see Sect. 3.4). In the latter case, we did not neglect the term $(\nabla^2{\bf B}){\bf B}$ in Eq. B.6, as in \cite{borrero2021}, and we changed Eq. B.7 accordingly. Once these changes were implemented, we kept the main iterative process detailed in \cite{borrero2021} (see Fig. 2), which we summarize briefly here for the sake of completeness. First, we inverted the data assuming HE. In this step ($i=0$), we get a first estimation of the temperature, ${\rm T}^{0}$, gas pressure, ${\rm P}^{0}_{\rm gas}$, ${\bf B}^{0}$, and $v_{\rm los}^{0}$ in the ($x$, $y$, $z$) volume. Then, we started an iterative process in which we: (1) solve the MHS equation so that we get an updated ${\rm P^{i+1}}_{\rm gas}$ that is now consistent with the ${\bf B}^{i}$ and $T^{i}$ inferred in the previous step (either the one assuming HE -- $i=0$ -- or the one from the previous iterative step). Yet, by doing so, it is not guaranteed that the new set of atmospheric parameters still fits the observed spectra, so a second step has to be taken. (2) We proceeded with a new inversion, in which we fixed the ${\rm P^{i+1}}_{\rm gas}$ obtained in step 1, obtaining a new ${\rm T}^{i+1}$, a new ${\bf B}^{i+1}$, and a $v_{\rm los}^{i+1}$ that goes to point 1 again. This iterative process ends when the observed data are successfully fitted as defined by a $\chi^{2}$ metric and the relative change in the gas pressure in two successive iteration steps is smaller than a certain threshold.

\end{appendix}

\end{document}